\documentclass[graybox]{svmult}

\usepackage{mathptmx}       
\usepackage{helvet}         
\usepackage{courier}        
\usepackage{type1cm}        
\usepackage{makeidx}         
\usepackage{graphicx}        
\usepackage{multicol}        
\usepackage[bottom]{footmisc}

\makeindex             

\begin{document}

\title*{Deep 3D Convolutional Neural Network for Automated Lung Cancer Diagnosis }
%
%
\author{Sumita Mishra\inst{1}\ \and
Naresh Kumar Chaudhary \inst{2}\ \and
Pallavi Asthana\inst{1}\ \and 
Anil Kumar\inst{1}\
}
\authorrunning{S. Mishra et al.}
%
\institute{
\textsuperscript{1} Amity School of Engineering and Technology, Amity University, Lucknow Campus , India
\\{smishra3@lko.amity.edu} \and
\\\textsuperscript{2}Dr. Ram Manohar Lohia Awadh University, Faizabad, U.P., India\\}

%
%
\maketitle
\abstract*{Each chapter should be preceded by an abstract (10--15 lines long) that summarizes the content. The abstract will appear \textit{online} at \url{www.SpringerLink.com} and be available with unrestricted access. This allows unregistered users to read the abstract as a teaser for the complete chapter. As a general rule the abstracts will not appear in the printed version of your book unless it is the style of your particular book or that of the series to which your book belongs.
Please use the 'starred' version of the new Springer \texttt{abstract} command for typesetting the text of the online abstracts (cf. source file of this chapter template \texttt{abstract}) and include them with the source files of your manuscript. Use the plain \texttt{abstract} command if the abstract is also to appear in the printed version of the book.}
\abstract{.  Computer Aided Diagnosis has  emerged as an indispensible technique for validating the opinion of radiologists in CT interpretation. This paper presents a deep 3D Convolutional Neural Network (CNN) architecture for automated CT scan-based lung cancer detection system. It utilizes three dimensional spatial information to  learn highly discriminative 3 dimensional features instead of 2D features like texture or geometric shape whick need to be generated manually.  The proposed deep learning method automatically extracts the 3D features on the basis of
spatio-temporal statistics.The developed model is end-to-end and is able to predict malignancy of each voxel for given input scan. Simulation results demonstrate the effectiveness of proposed 3D CNN network  for classification of lung nodule in-spite of limited computational capabilities.  
}    
\section{Introduction}
\label{sec:2}
Lung cancer is a prominent cause for cancer related deaths in India and most of these deaths may be prevented through periodic assessment of an individual 's risk of lung cancer .  The stages of lung cancer are indicated in the range from initial stage to fourth stage. In early stages cancer is limited to the lung. Lung Cancer spreads to other areas of the body like liver in advanced stages and patient outcome in such cases is not favorable. One of the primary characteristics of lung cancer in early stages is presence of abnormal tissue growth in lungs known as pulmonary nodule. Screening with the use of low-dose helical computed tomography(CT) may improve the mortality rates in high risk individuals by 20\% compared to conventional chest X Ray [1]. 
\begin{figure}[b]
\texttt{}
\includegraphics[width=0.9\linewidth]{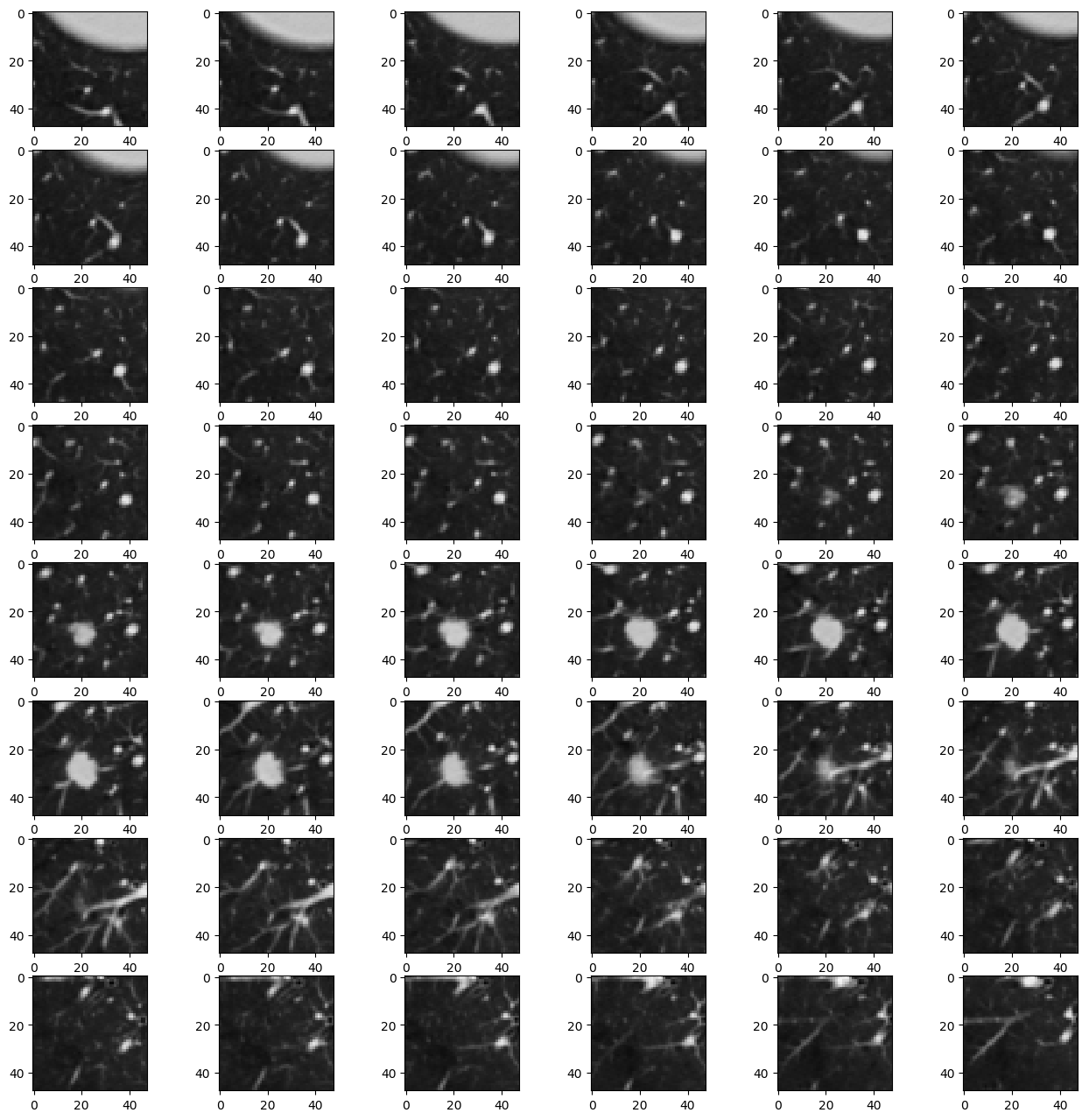}
%
%
\caption {A slice of CT Scan with malignant nodule from the dataset}
\end{figure}

In a conventional setup, the radiologist and oncologists play a crucial role in accurate diagnosis of Lung Cancer and it also depends upon the experience of specialist.In some cases even for highly trained radiologists, detecting nodules on CT scan and cancer diagnosis become challenging tasks.The recent progress in computer vision and deep learning may be exploited to achieve automatic detection of nodules to provide accurate clinical information and subsequently stage of progression of disease may be estimated. Deep Neural Networks (DNN) are being explored for developing automated tools for lung nodule detection because these algorithms are able to learn features from raw image data. DNN lung nodule identification methods can be classified into  2D deep CNN [2] and  3D deep CNN [3][4][5].

 The 2D CNNs treat each slice of the CT scan as an individual image, and training is done accordingly. However, in actual practice, nodules are dense 3D objects and in CT scans, they usually appear in several successive CT slices. Therefore, treating each slice as an individual image for nodule detection will result in the loss of highly correlative spatial context information.  In the 3D network, the convolutional kernels are 3 dimensional and the input data is 3D cubes instead of 2D images leading to increased accuracy, a major drawback of 3D CNN is increased complexity and requirement of more computational power.
 The proposed deep  3D convolutional neural network for pulmonary nodule classification  is capable of learning good 3 dimensional features without manual feature extraction and selection process required in conventional algorithms. This automatic 3D feature selection through the training of the network obtains precise characteristics of pulmonary nodules while retaining relatively better generalization capability of the network.

The paper is arranged in four sections. The next section deals with methods, Section III deals with Results and discussion, and section IV provides the concluding remark.

\section{Methods}
\label{sec:2}
 
The dataset used in the work is Luna which is based on publicly available Lung Image Database Consortium and Image Database Resource Initiative (LIDC/ IDRI) data[6]. CT scan Images in the database have nodule annotations from leading experts.  After data acquisition image preprocessing is performed  to make it suitable for further processing.

\subsection{Image Preprocessing}
\label{subsec:2}

The radius of the average malicious nodule in the dataset is 4.8 mm, with a typical CT volume of 400mm x 400mm x 400mm. First step in the process is data preparation the dataset is provided is in a medical imaging format called “MetaImage (mhd/raw)” which comprises header data in”. mhd” files and multidimensional image data is stored in “.raw files”. Three types of nodules are annotated in the dataset small nodules (nodule radius $<$ 3 mm), large nodules (nodule radius $>$ 3 mm) and non-nodules (non-cancerous benign nodules). The small nodules are represented only by center coordinate, Large nodules and benign nodules are annotated by a set of coordinates representing their edges. Non-nodules are benign nodules and therefore for training the network these are labeled as healthy tissue. After preprocessing malignant nodules are marked as positive samples and benign nodules are labeled as negative example. Further, different CT scans have different voxel length for the raw CT scans, so the next step is to convert these different CT scans to the same voxel spacing by resampling.  Sample-wide pixel normalization was applied afterwards to obtain uniform scaling between samples.  These resampled images are then saved  in numpy  format for further processing. Fig. 1 shows a typical input sample  having malignant  nodule.
\begin{figure}
\centering
\includegraphics[width=.95\linewidth]{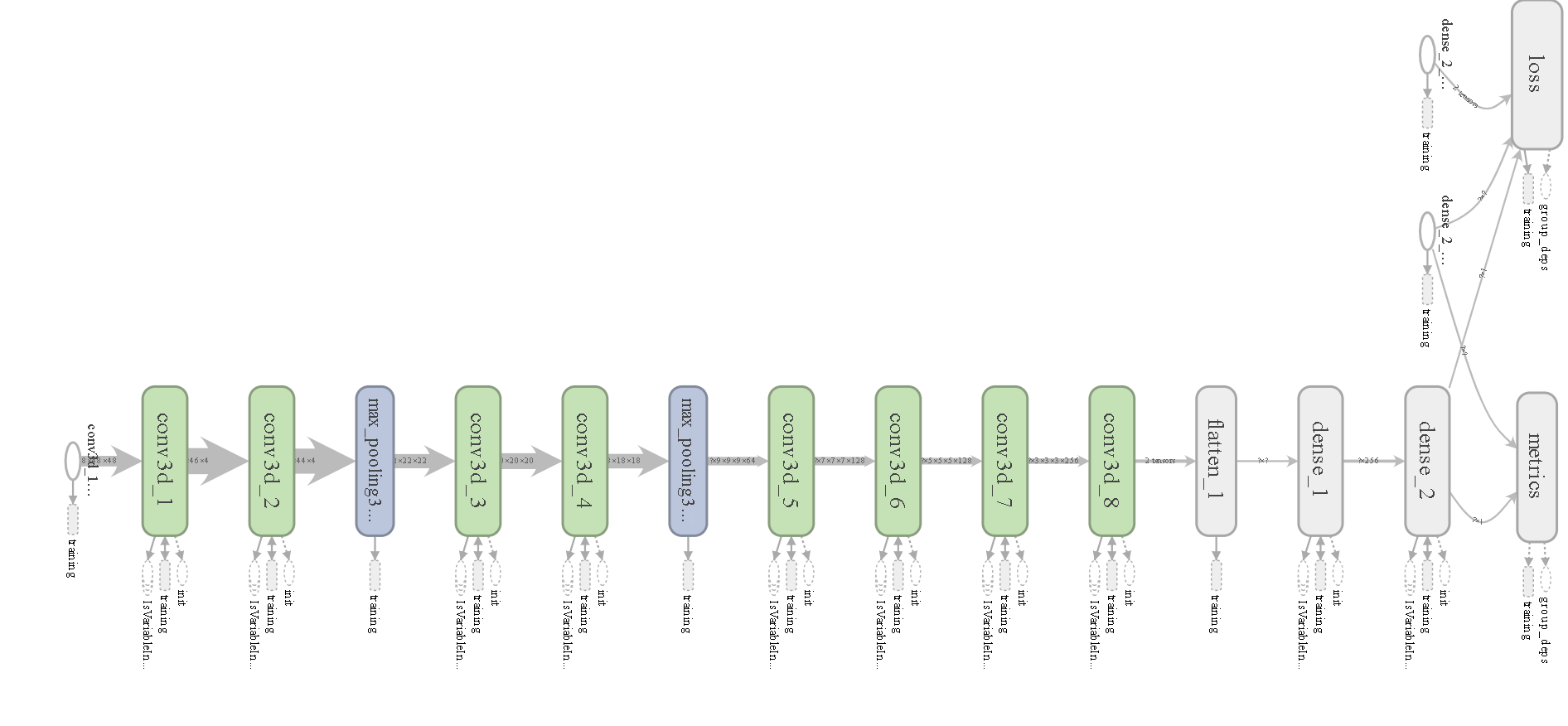}
%
%
\caption {Detailed Architecture of proposed Deep 3D CNN }
\label{2.png}       
\end{figure}

\subsection{	Architecture of Deep 3D CNN}
\label{subsec:2}

The detailed architecture of proposed deep, 3D CNN is shown in Fig. 2, it was implemented using keras API [7] and TensorFlow backend [8]  on CPU platform. Since Entire 3D CT Scan image cannot be fed as input to our 3D CNN model due to the CPU memory constraint; hence small 3D cubes are extracted from the lung scan volume and input to the network individually. Two kinds of cubes of size 48x48x48 are randomly selected. First set of the input is selected from positive samples so that they contain at least one malignant nodule. To ensure number of enough negative samples, second, set  of the inputs are cropped randomly from lung scans with random centers.  Latter category of inputs does not contain any malignant nodules. Thus, a single input sample to the network consists of 48x48x48 cube and the batch size of the input is 2. Gloret uniform initialization was used for all weights [9], with zero bias. Padding was applied around the edges of the input to preserve the dimensions of the data. We use two convolutional layers with 32  kernels followed by a max-pooling layer. Next two convolution layers have 64 kernels. These are followed by max-pooling layer. Two convolution layer with 128 kernals and two layers with 256 kernel are added followed by flatten and  two dense  layers. Kernel size in all the convolution layer is kept as 3x3x3 with stride of 1 in all three dimensions and all pooling layers have 2 x 2x 2 pool size. The output of the last layer is cancer probability of nodule which utilizes sigmoid activation function. 
The model was trained on 720  scans , 80 scans were used for validation of model and 88 scans were kept for evaluating the model on  test data. From each scan mini-batches were prepared and data was fed to the network in batch size of 2 due to limited memory capability.The  structure of each layer in the proposed network including input dimensions  is shown in Table 1.
Each convolutional layer uses Rectified Linear Unit (ReLU) nonlinearity for efficient gradient propagation.Although, ReLU may result in dead neurons but in our case, it led to fast convergence of data. Batch normalization was done to reduce saturation of neurons and improve generalization capabilities [10]. The average error of all pixels in an input sample defines the loss function which is an important metric to quantify the performance of the model on given sample. Overall performance of 3D CNN model is quantified by the average loss over all samples.We have employed Binary Cross Entropy loss function for training the model. Model is compiled using Stochastic Gradient Descent (SGD) optimizer with Nesterov momentum value of .9 [11],and learning rate of .003. This set of  parameters were found to accelerate the training process by faster convergence.  Stochastic gradient descent updates the weight matrix on the basis of negative gradient of the weights but momentum  causes the weights to also change in the direction of the previous update[12]. 

\begin{table}[h!]
  \begin{center}
    \caption{3D deep CNN Structure including Data Dimension.}
    \label{tab:table1}
    \begin{tabular}{l|c|r} 
     
     \hline
      \textbf{Layer (type)                 } & \textbf{Output Shape} & \textbf{Parameters}\\
      
      \hline
      conv3d1 (Conv3D)            & (None, 46, 46, 46, 32)    & 896       \\
     conv3d2 (Conv3D)     &       (None, 44, 44, 44, 32)   & 27680\\
     max-pooling3d1 (MaxPooling3)& (None, 22, 22, 22, 32) &   0         \\
     conv3d3 (Conv3D)     &       (None, 20, 20, 20, 64)  &  55360  \\
     conv3d4 (Conv3D)      &      (None, 18, 18, 18, 64) &   110656 \\   
     Max-pooling3d2 (MaxPooling3)& (None, 9, 9, 9, 64)&       0   \\      
     conv3d5 (Conv3D)      &      (None, 7, 7, 7, 128) &     221312    \\
     conv3d6 (Conv3D)      &      (None, 5, 5, 5, 128)  &    442496 \\   
     conv3d7 (Conv3D)      &      (None, 3, 3, 3, 256)   &   884992 \\   
     conv3d8 (Conv3D)       &     (None, 1, 1, 1, 256)    &  1769728 \\  
     flatten1 (Flatten)      &    (None, 256)          &     0         
     \\
     dense1 (Dense)           &   (None, 256)          &     65792     
    \\
     dense2 (Dense)            &  (None, 1)             &    257       
     \\
     \hline
     
     \multicolumn{3}{c}{{Total Trainable parameters: 3,579,169}}  \\
     \hline
    \end{tabular}
  \end{center}
\end{table}

\section{Results}
\label{sec:3}
\begin{figure}[!tbp]
  \centering
  \begin{minipage}[b]{0.4\textwidth}
    \includegraphics[width=\textwidth]{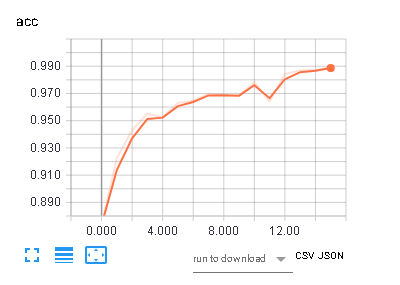}
    \caption{Training Accuracy during training.}
  \end{minipage}
  \hfill
  \begin{minipage}[b]{0.4\textwidth}
      \includegraphics[width=\textwidth]{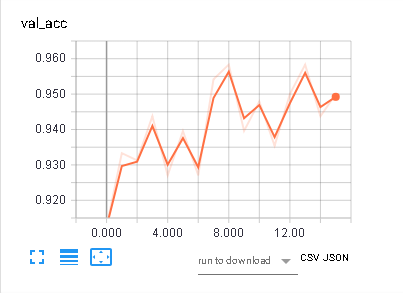}
      \caption{Validation Accuracy during training.}
    \end{minipage}
  \begin{minipage}[b]{0.4\textwidth}
    \includegraphics[width=\textwidth]{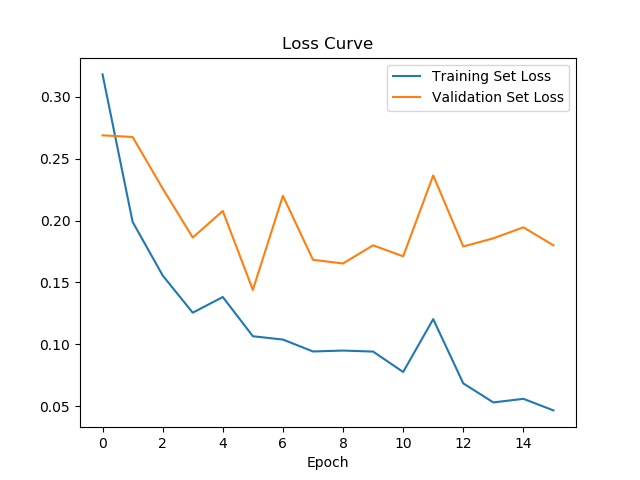}
    \caption{Loss Curve.}
  \end{minipage}
\end{figure}
The proposed 3D CNN architecture achieved an accuracy of 94.8\% on validation data. Figure 3 shows the training accuracy and validation accuracy as the model was trained. It may be observed from the figure that the accuracy is gradually increasing during training. Figure 4 shows that the loss curve of training samples and validation samples.To improve the quality of this model, the validation loss and validation accuracy  over epochs was monitored using tensorboard logs and  hyper parameters were appropriately adjusted.
 The performance with respect to accuracy becomes stable after several iterations. This behavior is correlated with the change in loss as shown in Fig 5 which is decreasing gradually during training, and when the network obtains an optimal point then the training process becomes stable. Overall, the proposed deep 3D convolutional neural network is effective  in classification as shown in fig 4 depicting the accuracy during  the validation process.
In order  to demonstrate the performance of our model a complete CT scan containing annotated malignant nodule was sampled into multiple cubes using sliding window and subsequently fed to the trained 3D CNN network and  probabilistic predictions corresponding to each input sample is plotted in 2 Dimension. Fig 6 shows the predicted probability map from the model in 2 Dimension while fig 7 shows the regions with probability greater than 90\% after removal of noisy predictions .Fig 8 shows 2D slice of corresponding input sample. It may be observed  that  location of the annotated nodules has very high-probability but a small number of false positives are also predicted as shown in Fig 6. Evaluation of model on test data yielded a score .208 and accuracy of .9514.Due to the extensive duration of the training process hyper-parameters initial learning rate, momentum decay was adjusted only once. Further optimization of these parameters  is expected to improve the quality of the model and reduce false positives.

\begin{figure}[!tbp]
  \centering
 
  \begin{minipage}[b]{0.48\textwidth}
    \includegraphics[width=\textwidth]{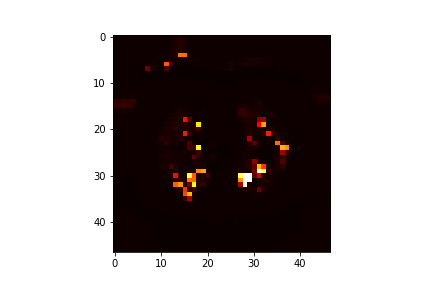}
    \caption{Prediction of probability map by model.}
  \end{minipage}
 \hfill
  \begin{minipage}[b]{0.48\textwidth}
        \includegraphics[width=\textwidth]{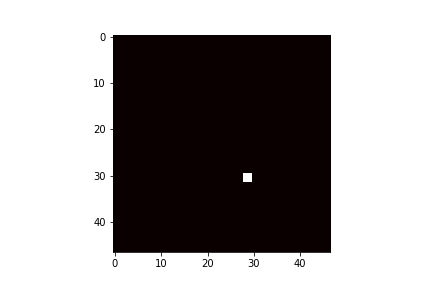}
        \caption{ predicted sample result.}
      \end{minipage}
 \begin{minipage}[b]
    {0.38\textwidth} 
       \includegraphics[width=\textwidth]{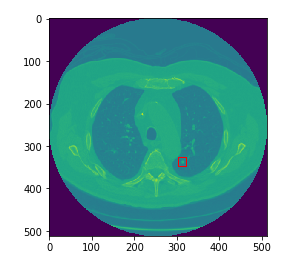}
       \caption{Malignant nodule.}
     \end{minipage}
  
\end{figure}
Table 2 shows the comparison of our work with similar works and it must be noted  that  different authors use different re-sampling and data division methodology for generating test train and validation data. In addition,only some of these methods are end-to-end thus making  it difficult to achieve a fair comparison.
\begin{table}[h!]
  \begin{center}
    \caption{Performance Comparison of Proposed work with Other Published Work}
    \label{tab:table1}
    \begin{tabular}{l|c|r} 
     
     \hline
      \textbf{Related work             } & \textbf{Dataset characteristics} & \textbf{Performance }\\
      \hline

     D. kumar et al  2015  [13]   &       LDRI, nodule size $>$  3mm  &  Accuracy 75.01\%  \\
      
     Gruetzemacher, R. et al  2016  [14]      &      LDRI both small and large nodules  &    Accuracy 86.13\% \\

    Hamidian s. et al   2017 [15]      &      nodule size $>$, 3mm &      Sensitivity   95\%  \\
    
     Wu et al   2017 [16]             &  NCI, China  ( nodule size not specified  )   
       
       &    Accuracy 77.8\%     
     \\
      
      Proposed work   &       LDRI, nodule size $>$  3mm  &  Accuracy 94.80\%\\
     \hline

     \hline
    \end{tabular}
  \end{center}
\end{table}

\section{Conclusion}
\label{sec:4}
The chances of successful outcome of lung cancer treatment are improved if it is detected in early stages.  In this paper, a deep 3D convolutional neural networks is presented  to make predictions regarding the presence or absence of a malignant pulmonary nodule in CT scan. The  proposed 3D Deep CNN automatically extracts 3D features for pulmonary nodule classification directly from CT volume.We have demonstrated that our 3D Deep CNN model is able to perform fairly and  achieves an accuracy of 94.8\% without extensive pre-processing and limited computational capabilities .The performance of the proposed model may be further improved by utilizing suitable data augmentation technique. Although data augmentation using axes swapping was tried in this work but no significant improvement is obtained and over-fitting may be reduced by using drop out layer.





\end{document}